\documentclass[final,5p,times,twocolumn]{elsarticle}

\usepackage{graphicx}
\usepackage{amssymb}
\usepackage{physics}
\usepackage{enumitem}
\usepackage{subfig}
\usepackage[]{algorithm2e}
\usepackage{animate}

\usepackage{lineno}

\def\fig{Figure}

\def\Fig{Figure}

\def\sect{Section}

\def\tab{Table}

\def\Tab{Table}

\def\eqn{equation}

\def\fitswarp{{\sc Fits\_Warp}}

\journal{Astronomy and Computing}

\begin{document}

\begin{frontmatter}


\title{De-distorting ionospheric effects in the image plane}


\author[CIRA]{N. Hurley-Walker}
\author[CIRA]{P. J. Hancock}
\address[CIRA]{International Centre for Radio Astronomy Research, Curtin University, Bentley, WA 6102, Australia}

\begin{abstract}
The Earth's ionosphere refracts radio waves incident on an interferometer, resulting in shifts to the measured positions of radio sources. We present a method to smoothly remove these shifts and restore sources to their reference positions, in both the catalogue and image domains. The method is applicable to instruments and ionospheric weather such that all antennas see the same ionosphere. The method is generalisable to repairing any sparsely-sampled vector field distortion to some input data. The code is available under the Academic Free License\footnote{https://opensource.org/licenses/AFL-3.0} from \texttt{https://github.com/nhurleywalker/fits\_warp}.
\end{abstract}


\end{frontmatter}


\section{Introduction}\label{sec:intro}

In recent years there has been a resurgence in low-frequency radio observing, in part due to endeavors to detect the Epoch of Reionisation via its redshifted 21-cm emission. Covering frequencies between 30 and 300\,MHz, low-frequency telescopes built in the last decade include the Long Wavelength Array (LWA; \citealt{Taylor}), the Low-Frequency Array (LOFAR; \citealt{vanHaarlem}) and the Murchison Widefield Array (MWA; \citealt{Tingay13, Lonsdale}). Construction of the low-frequency component of the Square Kilometer Array is imminent. These telescopes make use of an interferometric design, in which the signals from multiple antennas are correlated together to produce a set of ``visibilities'', which is a sampled Fourier Transform of the sky.

\subsection{The problem of ionospheric distortion}

When performing imaging operations with these telescopes, a common issue faced by observers is that of ionospheric distortions to the incident radio waves from celestial radio sources. The Earth's ionosphere consists of layers of partly-ionised plasma at altitudes from around 60 to 1,000\,km. Its electron density varies with altitude and time of day and ranges between $10^4$ and $10^6$\,cm$^{-3}$. As such it acts as a refractive medium for incident radio waves, with line-of-sight refractive shifts proportional to the square of the wavelength of the incident wave.

The total electron column density (``total electron content'' (TEC)) adds an equal phase to all interferometric antennas. Interferometers measure angular positions using phase differences between antennas, and so are insensitive to this constant offset component. Instead, the transverse gradient $\nabla_{\perp}$ in the TEC toward the source introduces an angular shift $\Delta \theta$ in the position of a radio source, which is given by \cite{TMS}:

\begin{equation}
\Delta \theta = - \frac{1}{8\pi^2}\frac{e^2}{\eta_0 m_e } \frac{1}{\nu^2} \nabla_\perp \mathrm{TEC}
\label{eq:d_theta}
\end{equation}

$e$ and $m_e$ are the electron charge and mass, $\eta_0$ is the vacuum permittivity, and $\nu$ is the radio observing frequency. The negative sign indicates that the direction of refraction is toward decreasing TEC.

\citet{2005ASPC..345..399L} explore some of the considerations for designing low-frequency radio telescopes in this regime. 
\fig~1 of that paper shows a schematic overview of the different conditions that may be faced by these arrays: a telescope may have short baselines, in which case all antennas see the same $\nabla_\perp \mathrm{TEC}$ along a particular line-of-sight, or it may have long baselines, in which case antennas could see a different $\nabla_\perp \mathrm{TEC}$.
A telescope with a narrow field-of-view will only see a single $\nabla_\perp \mathrm{TEC}$ and thus need a single phase correction over the field.
Whereas, a telescope with a wide field-of-view can possibly see multiple different $\nabla_\perp \mathrm{TEC}$ and therefore need multiple phase corrections over the field.
Since the field of view is set by the individual antennae and the baseline length is set by the telescope layout, telescopes can conceivably be built with any combination of baseline length and field-of-view.
Interferometers with a wide field-of-view and short baselines require a direction dependent phase correction, but can use the same phase correction for all antennae.
If such an instrument can be calibrated, the direction dependent phase corrections can be applied in the image domain by warping the resulting images, thus undoing the position shifts described by \eqn~\ref{eq:d_theta}.

A critical quantity for the calibration of an interferometer is the diffractive scale size, $r_\mathrm{diff}$, which is the ionospheric patch size over which the phase difference due to changes in $\mathrm{TEC}$ is less than $\pi$\,rad, compared to the longest baseline of the telescope (\citet{2016RaSc...51..927M}).
The variance in phase $\phi$, seen on a baseline of length $r$, in the presence of power-law ionospheric turbulence, is: 
\begin{equation}
D(r) = \langle\left( \phi(r') - \phi(r'+r)\right)^2\rangle = \left(\frac{r}{r_\mathrm{diff}}\right)^\beta
\label{eq:dr}
\end{equation}

When $r_\mathrm{diff}$ becomes small the phase variance on long baselines becomes large and it becomes necessary to derive baseline-dependent calibration solutions.
A small $r_\mathrm{diff}$ also corresponds to a shorter timescale for the phase variance and thus the calibration solutions need to be derived at a higher cadence, quickly reducing the signal to noise (and availability) of suitable calibrator sources.
When $r_\mathrm{diff}$ is large compared to the longest baseline, the phase variance between antennae becomes small as they effectively see the same ionosphere.
In such cases a solely antenna-based gain calibration can be calculated and applied, often without the need for any time dependence.
Interferometers with a wide ($\approx30^\circ$) field-of-view and relatively short baselines, such as the MWA, may sample the ionosphere every $\approx1$--$20$\,km, over projected areas of $\approx15$--$300$\,km, depending on altitude, allowing relatively easy calculation of the critical $r_\mathrm{diff}\sim 4$\,km.
\citet{2017MNRAS.471.3974J} made extensive investigations into typical ionospheric behaviour above the MWA and estimated that $r_\mathrm{diff}$ varies from $\approx3$--8\,km above the MWA, compared to its original maximum baseline length of 2.5\,km.
Over their 19~nights of observing, they found that $r_\mathrm{diff}<4$\,km occurs only $\approx10\%$ of the time.
\citet{GLEAMEGC} found a similar result, observing 30 nights and only discarding two due to poor ionospheric conditions indicative of low $r_\mathrm{diff}$.
In the Northern hemisphere, \citet{2016RaSc...51..927M} draw similar conclusions from LOFAR data: $r_\mathrm{diff}<5$\,km for about 10\% of their 29 nights of observing.

For those observations with wide fields-of-view, and $r_\mathrm{diff}$ larger than the longest baseline of the interferometer, the resulting image of the sky still contains phase variations across the image, which can be seen via the $\Delta \theta$ of the individual radio sources.
As \citet{2016RaSc...51..927M} points out, these large scale variations in $\nabla_\perp \mathrm{TEC}$ are not part of the power-law turbulence described by \eqn~\ref{eq:dr}, but are due to coherent structures in the ionosphere such as traveling ionospheric disturbances, or field aligned plasma tubes \citep{2015GeoRL..42.3707L}.
The $\Delta \theta$ are essentially a foreground effect which for most astronomical purposes needs to be modeled and removed, for instance so that association with sources from other astronomical catalogues can be accurately performed, or to successfully combine multiple observations  without blurring the resulting effective resolution element, or ``point spread function''.

\subsection{Existing solutions}

The optical and infrared astronomy community face similar challenges: the alignment of charge-coupled devices (CCDs) may not be precisely known, introducing an instrumental shift to the position of the detected sources, and for ground-based telescopes, the troposphere may refract and scintillate incident wavefronts. The timescales of the latter distortion are so short ($\approx$milliseconds) that the real-time hardware solution of adaptive optics must be used for optimum imaging fidelity (see \citet{2012ARA&A..50..305D} for a review). 

While the first problem bears some similarity to the ionospheric distortion of radio images, the instrumental shifts are usually fairly simple in form, e.g. a set of linear translations, scale changes, and rotations. Solutions such as the MoSaic data REDuction (\textsc{mscred}) tool in the Image Reduction and Analysis Facility \citep[\textsc{IRAF};][]{1986SPIE..627..733T,1993ASPC...52..173T}\footnote{http://iraf.noao.edu/} package, or Software for Calibrating AstroMetry and Photometry \citep[\textsc{SCAMP}; ][]{2006ASPC..351..112B} in the \textsc{Astromatic}\footnote{http://www.astromatic.net/} ecosystem allow the user to match detected source positions with objects with known high-precision positions, such as stars, and then calculate the resulting transforms that need to be applied, which are usually saved to the \textsc{FITS} header rather than applied to the image data itself. Unfortunately, no combination of these transforms is adequate to describe the complex ionospheric distortions, and there is the added complication that the image projections used by optical astronomers (such as Distorted Tangential; TPV) are not optimal for the extremely wide field-of-view of telescopes like the MWA, requiring reprojections that would distort the radio images and add complication to the calculation of the point spread function.

Correcting radio astronomy data for ionospheric distortions is not a new problem, but the regime faced by the MWA is substantially different from previous experiments and requires a different approach. For example, \citet{2005ASPC..345..337C} use a field-based calibration technique to improve image fidelity of the Very Large Array in its long-baseline “A” (baseline lengths $\leq36.4$\,km) and “B” (baseline lengths $\leq11.1$\,km) configurations. This technique finds a phase correction for each antenna toward a bright source in the field-of-view, fits the ionospheric phase gradient over the telescope array, and then applies that model during the imaging process. This technique works for the VLA due to the very high sensitivity of each individual antenna, which allows the gain solutions to be calculated every thirty seconds for every antenna. The Source Peeling and Atmospheric Modeling \citep[SPAM; ][]{2009A&A...501.1185I} improves on this by using 10--20 sources in the field-of-view, and is typically used by the Giant Metrewave Radio Telescope \citep[GMRT; ][]{1991CuSc...60...95S} to form high-fidelity low-frequency images.

In comparison to the VLA and the GMRT, the MWA has a field-of-view an order of magnitude larger, and a $\approx50\times$ lower antenna sensitivity. Solving for per-antenna phases toward all bright sources in the field-of-view over short temporal cadences to a sufficiently accurate level to model the ionosphere is challenging because of the comparably lower signal-to-noise on each source. \citep{2016MNRAS.458.1057O} use clusters of sources to build up signal-to-noise on each cluster, solving for the per-antenna gains and peeling the clusters of sources from the data. However, this technique currently only works at high elevations in the middle of the MWA band, where the telescope is most sensitive.

These visibility-based techniques are not necessary for the ionospheric regime where all antennas of the array view the same ionosphere, i.e. the array is not defocussed. In this situation, we can attempt an image-based solution, which is the focus of this work.

\subsection{This work}

\citet{GLEAMEGC} briefly introduce a method to perform an image-based correction; this work provides a more extensive explanation of the technique (\sect~\ref{sec:method}) including details of the implementation, tests on the effectiveness and robustness of the method (\sect~\ref{sec:results}), and concludes with thoughts on future uses of this code (\sect~\ref{sec:conclusions}).

\section{Method}\label{sec:method}
The algorithm, entitled \fitswarp{}, performs two tasks: cross-matching catalogues, and warping images.
The image warping is done using a model of pixel offsets that are derived from a catalogue of cross-matched sources.

The observation for which the correction will be performed needs to be either an image in the format Flexible Image Transport System (\textsc{fits}), with a correct World Co-ordinate System (\textsc{wcs}) header, or a catalogue of sources with approximately correct celestial positions. In either case, an initial cross-match with a correct reference catalogue must give more correct matches than false; accuracy to a few arcmin should be adequate at MWA frequencies. The number of radio sources visible needs to be sufficient that the field-of-view is well-sampled without being confused; in the MWA regime of ionospheric distortions, this is at least one source per square degree.

The catalogue of cross-matched sources can be created using an external program, or by \fitswarp{} itself.
Here we discuss the two methods.

\subsection{Cross-matching catalogues}
The goal of this stage is to cross-match a reference catalogue with a target catalogue.
The reference catalogue is assumed to have a good (correct) astrometry solution, whilst the target catalogue has only an approximate astrometry.
The input and target catalogues are assumed to cover the same region of sky, with the input catalogue being cropped as required.
Cross-matching is performed in two phases using an iterative approach.
The first stage aims to correct for a bulk offset between the catalogues.
The average or center coordinates for the reference catalogue are calculated.
A spherical rotation is applied to both the target and reference catalogues so that they have lat/lon measured relative to the center coordinates calculated above.
The two catalogues are cross matched using a standard approach\footnote{e.g. Topcat \citep{Taylor_topcat_2005} or the \texttt{coordinates} module of Astropy \citep{TheAstropyCollaboration2013,2018arXiv180102634T}} to calculate the a bulk lat/lon offset between the two.
This bulk offset is applied to the target catalogue.

The second stage aims to correct sub-image-scale distortions.
Ionospheric distortions have coherent features on angular scales larger than about $1^\circ$. They can be persistent for minutes to hours, or change rapidly over tens of seconds. Some ionospheric structures can lead to persistent features with a parameterizable form, such as the appearance of sinusoidal peaks and troughs of $\Delta\theta$ generated by Whistler ducts \citep{2015GeoRL..42.3707L}. However, in general, the distortions are not easily parameterized by simple mathematical functions. We therefore crossmatch the catalogues again, and fit the lat/lon offsets using an ensemble of radial basis functions with a simple linear form, as implemented in the \textsc{scipy}\citep{Jones_scipy_2001} function \texttt{interpolate.Rbf} \footnote{https://docs.scipy.org/doc/scipy/reference/generated/scipy.interpolate.Rbf.html}.

\begin{figure*}
\centering
\subfloat[all sources\label{subfig:sep_cat_all}]{\includegraphics[width=0.5\textwidth]{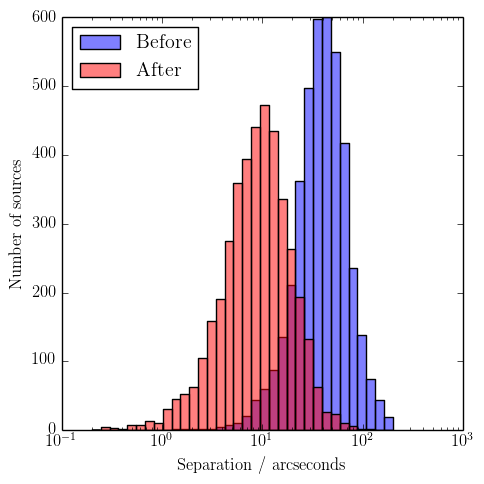}}
\subfloat[sources with S/N$>20$\label{subfig:sep_cat_bright}]{\includegraphics[width=0.5\textwidth]{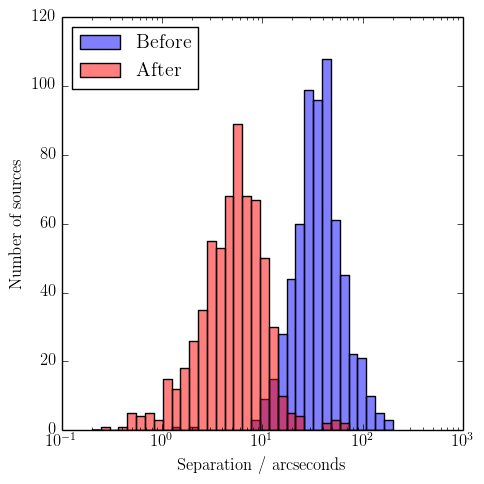}}
\caption{Separation between target and reference source positions in arcseconds, along great circles, before (blue) and after (red) applying \fitswarp{} to the catalogue.
The first panel shows all sources and the second panel shows only the bright (S/N$>20$) sources. The abscissa is a log scale. By applying the algorithm, we can measure a factor of 3 improvement in separation for all sources, rising to a factor of 10 for bright sources.\label{fig:xmatch}}
\end{figure*}

The RBF implementation allows for a smoothing factor to be implemented, which is used to mitigate the effects of false matches, and to provide a spatial averaging to the calculated offsets.
The smoothed lat/lon models are then evaluated at each of the target positions, and the target catalogue is updated accordingly.
The crossmatch-model-update loop is performed three times.
The third and final stage is then to crossmatch the target and reference catalogues, and return a joined catalogue using the initial (uncorrected) target positions in the initial (prior to rotation) coordinate frame. \Fig~\ref{fig:xmatch} shows the separations of detected sources from their reference positions before and after correction.

\subsection{Correcting images}

Processing radio interferometer visibility data into clean, calibrated images of the sky is extensively discussed elsewhere; see \cite{TMS} for an overview. The  important step for this discussion is that the visibilities are Fourier Transformed into the image plane, and the result is a projection of the celestial sphere on to a plane that is tangent at the field centre. This is an oblique orthographic projection, with the pole at the field centre of the observation, commonly denoted as a \textsc{SIN} projection.

The basis frame for ionospheric distortions is the physical sky, rather than celestial co-ordinate frames. Thus all measurements and corrections for these distortions should be performed in a sky frame rather than a celestial frame. Conveniently for this work, this means that the $(x,y)$ pixel frame of the \textsc{SIN}-projected images is a good natural frame in which to work, rather than any celestial co-ordinate system derived from the observatory position and time of observation. 

The astronomer also needs a reference catalogue with astronomically-correct source positions and a source density at least comparable to that of the input image, in order for cross-matching to be possible. Best performance will be realised if the input and reference catalogues are both dense enough to sample expected ionospheric features, which are of scales $\approx1^\circ$--$10^\circ$. Performance may become slower if the catalogue and image oversample the ionosphere such that the functions must be fit over a large ($\gtrsim$1,000) number of points.

At low ($<1$\,GHz) frequencies where ionospheric effects are important, a wide variety of surveys are available, depending on desired frequency and sky position. For the work presented in this paper, we use the first extragalactic catalogue \citep{GLEAMEGC} from the GaLactic and Extragalactic All-sky Murchison Widefield Array survey\citep[GLEAM;][]{2015PASA...32...25W}, which has a source density of $\approx20$ sources per square degree, i.e. more than sufficient to sample ionospheric features, and an astrometric accuracy of $\approx1$" over most of the sky.

The method then proceeds as follows:
\begin{enumerate}[label=\textbf{Step~\arabic*}]
\item{Calculate the $(x,y)$ offsets for each source, $(\Delta x,\Delta y)$;}
\item{Use radial basis functions to fit a general non-parametric model to the vector offsets $(\Delta x$,$\Delta y)$;}\label{fitrbf}
\item{For each pixel in the original image, with flux value $F$, calculate the corrected position $(x+\Delta x$, $y+\Delta y)$, from the model;}\label{callmodel}
\item{Interpolate over the ensemble of $(x+\Delta x$, $y+\Delta y)$ positions and their fluxes $F$ to calculate the modified flux value, $F'$, at the original pixel positions $(x$, $y)$;}\label{interpolate}
\end{enumerate}

\section{Testing}\label{sec:results}

We wish to test two main attributes of the algorithm:
\begin{enumerate}[label=\textbf{Test~\arabic*}]
\item{Does \fitswarp{} correctly remove the warping effect of the ionosphere, bringing the measured source positions in line with those in the reference catalogue?}\label{test1}
\item{Does \fitswarp{} preserve image fidelity and source attributes, such that the measurements made on the warped image are still useful for astronomical purposes?}\label{test2}
\end{enumerate}

To test these attributes we perform testing on real, partly-simulated, and fully-simulated data. In order to measure the positions of sources in images in these tests, we use the source-finding algorithm \textsc{Aegean} \citep{Hancock_compact_2012,2018PASA...35...11H} and its associated Background And Noise Estimator (BANE). These programs are optimized for detecting and characterizing isolated Gaussian sources typically found in extragalactic radio images.

An important correlate in this testing is the divergence of the vector fields, since the larger the divergence at a given location, the more interpolation is being performed. We therefore write out divergence maps for every image, i.e. $\dv{\Delta x}{x} + \dv{\Delta y}{y}$ at every pixel location.

\subsection{Testing on MWA data}\label{sec:realdata}

In order to test \fitswarp{} we identified data in the MWA archive\footnote{https://asvo.mwatelescope.org} which was known to have ionospheric distortions when imaged.
Twenty such observations were identified and are listed in \tab~\ref{tab:obsids}.
The observations were taken between 2013-08-27 to 2014-11-03, and all were observed at a central frequency of 119\,MHz (with contiguous bandwidth 30.72\,MHz), with a common pointing center of RA=42$^\circ$, Dec=-26.5$^\circ$, for a duration of 112\,s.
The observations were downloaded from the MWA archive, calibrated using a nearby observation of Pictor A, and imaged using \textsc{WSClean} \citep{2014MNRAS.444..606O}. The pixel resolution of each resulting image is 42.58~arcsec and the synthesised (restoring) beam is circular with FWHM 2.68~arcmin.

This calibration and imaging process uses a direction- and baseline-independent calibration solution, resulting in an image where the sources themselves are not distorted (i.e. $r_\mathrm{diff}>2.5$\,km), but which has phase variations across the field of view. 
These phase variations manifest as source position shifts and thus the image can be corrected using \fitswarp{}.

We apply the \fitswarp{} algorithm to these images, using the GLEAM catalogue as the reference catalogue for the positions of sources. The resulting vector distortions and the RBF model (sampled over a grid of $100\times100$ pixels) are shown in \fig~\ref{fig:vector_field}. We note that this figure shows that the method is robust to the occasional incorrect crossmatch (single vectors that do not match the general trend of the vector field).

We then perform source-finding on the resulting warped image, and compare the separation of sources from the reference catalogue before and after warping (\ref{test1}). To perform \ref{test2}, we measure any differences in source properties, i.e. integrated and peak flux density, source shape and size (major and minor axis lengths), and also re-measure the noise using \textsc{BANE} to determine whether the noise properties have changed.

\begin{figure*}
\centering
\animategraphics[controls,width=\textwidth,poster=first]{1}{lowres/image_}{01}{21} 
\caption{An animation, at one frame per second, iterating through the twenty observations listed in \Tab~\ref{tab:obsids}. The left panel shows the measured vector offsets of the sources from their reference positions. The vector lengths are in pixels rather than degrees, i.e. magnified by a factor of 85 (the inverse of the pixel size in degrees). The right panel shows the RBF model sampled over a grid of $100\times100$ pixels. The colourbar shows the angle of the vectors counterclockwise from West, in degrees. The first frame shows observation 1068480760, which is discussed in more detail in the text.\label{fig:vector_field}}
\end{figure*}

\begin{figure*}
\centering
\subfloat[all sources\label{subfig:sep_all}]{\includegraphics[width=0.5\textwidth]{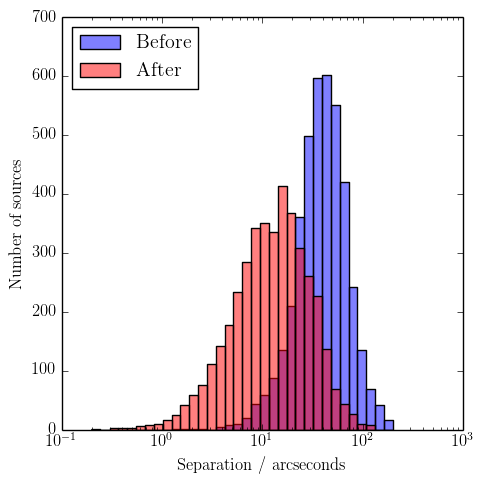}}
\subfloat[sources with S/N$>20$\label{subfig:sep_bright}]{\includegraphics[width=0.5\textwidth]{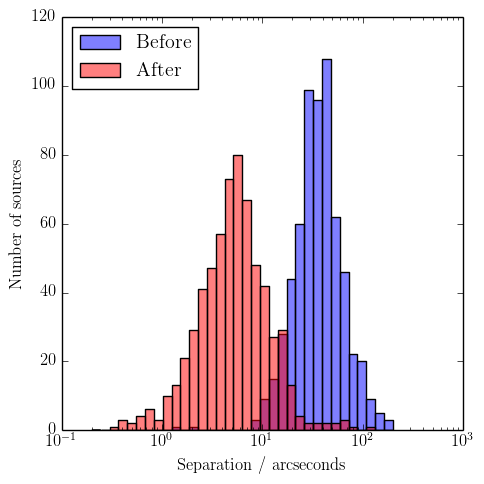}}
\caption{Separations of sources in arcseconds, along great circles, before (blue) and after (red) applying \fitswarp{}. The first panel shows all sources and the second panel shows only the bright (S/N$>20$) sources. The abscissa is a log scale. By applying the algorithm, we can measure a factor of 3 improvement in separation for all sources, rising to a factor of 10 for bright sources.\label{fig:seps}}
\end{figure*}

\fig~\ref{fig:seps} shows histograms of the separations of the sources from the reference catalogue before and after warping. From this we can see that \fitswarp{} moves the sources much closer to the positions listed in the reference catalogue. Given that the resolution of the image is 2.68', we can predict a typical position accuracy of 4" for sources with S/N$>20$ \citep[Equation 14-5 of ][]{1999ASPC..180..301F}. This compares favourably to the median final separation of 5" for these sources, after applying \fitswarp{}. \fig~\ref{fig:seps} closely resembles \fig~\ref{fig:xmatch}, showing that the correction works similarly in catalogue and image space, as it should.

Having successfully passed \ref{test1}, we turn to \ref{test2}. The noise results are straightforward: BANE measures direction-independent differences of $\approx10\mu$Jy\,beam$^{-1}$ between the unwarped and warped images, compared to typical noise levels of $\approx35$mJy\,beam$^{-1}$, i.e. more than 3 orders of magnitude smaller. We therefore believe the noise characteristics to be essentially unchanged by \fitswarp{}.

\begin{figure*}
\centering
\subfloat[Integrated flux density ratio (after / before)\label{subfig:int_flux_real}]{\includegraphics[width=0.5\textwidth]{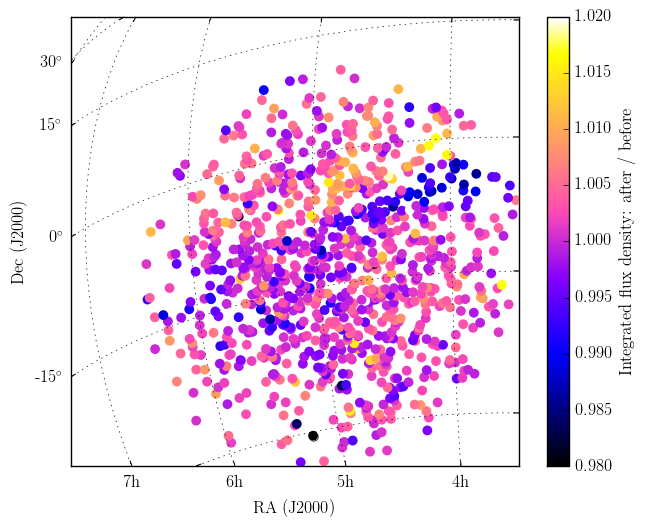}}
\subfloat[Divergence\label{subfig:div_real}]{\includegraphics[width=0.5\textwidth]{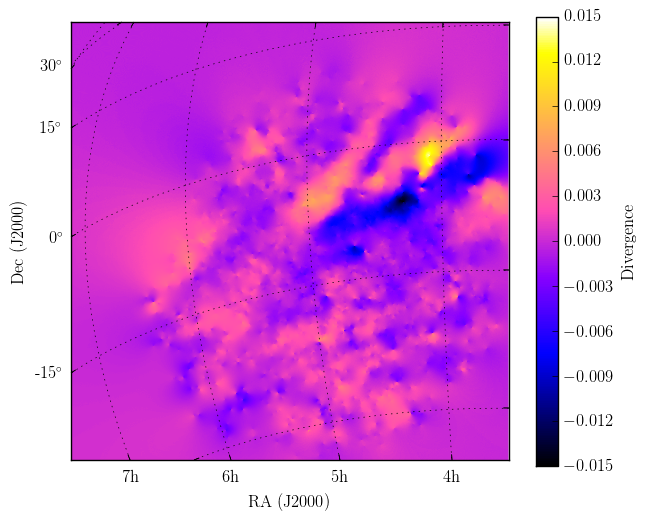}}
\caption{For observation 1068480760, the integrated flux density ratio after:before applying \fitswarp{} (left) and the divergence map of the vector field (right). In the left panel, only sources with S/N$>15$ are used. A correlation can be seen between the integrated flux density ratio and the divergence map in the regions of particularly strong divergence near the top right.\label{fig:int_div_real}}
\end{figure*}

Much examination of the source parameters was performed and little deviation detected, but the fainter sources are not useful for this test and added a considerable amount of noise to the results. Also, the distribution of sources over the sky is semi-random and not all areas are equally-sampled. We therefore show only the most visible result, which is a slight distortion to the integrated flux density of sources in an area of large divergence. In order to demonstrate the maximum effect, we plot the integrated flux density ratio and divergence map of the observation, which has a particularly coherent ionospheric feature with large ($\approx 2\%$) divergence (\fig~\ref{fig:int_div_real}). We cannot use these data alone to quantify whether \fitswarp{} passes \ref{test2}. Therefore, we turn to partially simulating the data in order to more thoroughly test the algorithm.

\subsection{Testing on partially-simulated data}

For every snapshot, we replace the image data with a grid of $\approx$17,000 Gaussian sources (major and minor axes $=2.68$'), with peak flux densities of 2.123\,Jy. They are separated by 30\,pixels ($=21.75$') in each ($x$,$y$) direction. We re-simulate the noise to be constant across the image, at a level of 16.5\,mJy\,beam$^{-1}$, giving every source a signal-to-noise ratio of $\approx130$. We retain the cross-matched catalogues from Section~\ref{sec:realdata} and use them as the input to the determine the warping fields, then apply each (unique) warp to the (identical) simulated images.

We source-find on each warped image and compare the resulting source parameters to those of sources detected in the original image. To avoid areas where the vector field is entirely extrapolated, which would normally correspond to areas where the telescope has no sensitivity, we restrict further analysis to sources within $20^\circ$ of the pointing centre.

\begin{figure*}
\centering
\subfloat[major axis, $a$\label{subfig:a}]{\includegraphics[width=0.4\textwidth]{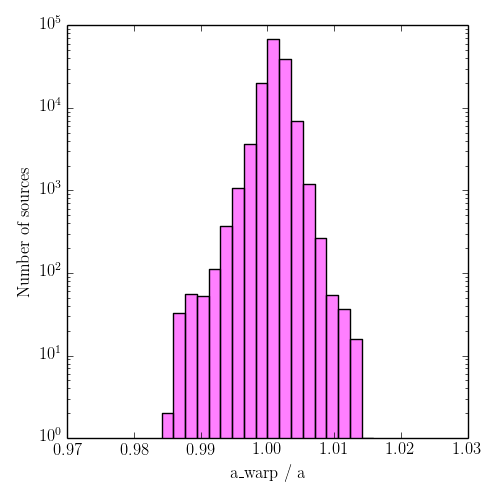}}
\subfloat[minor axis, $b$\label{subfig:b}]{\includegraphics[width=0.4\textwidth]{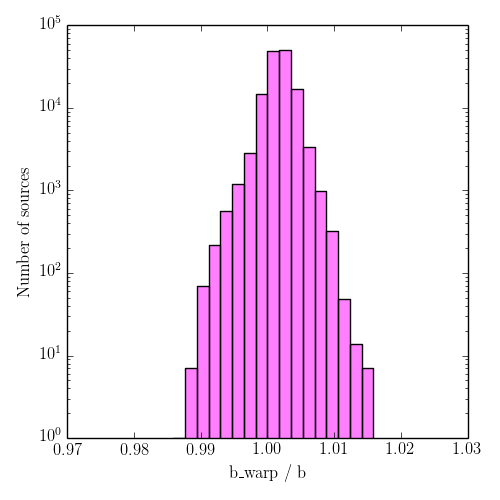}}

\subfloat[integrated flux density\label{subfig:int}]{\includegraphics[width=0.4\textwidth]{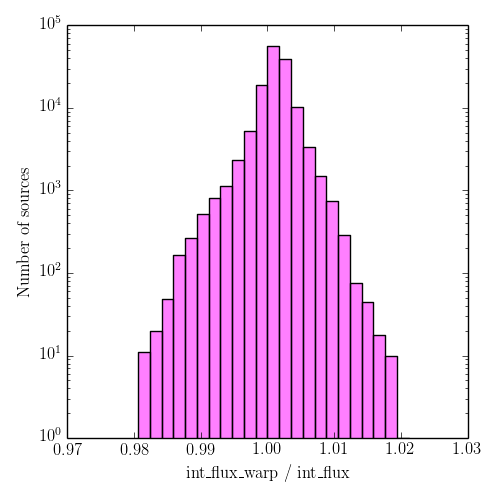}}
\subfloat[peak flux density\label{subfig:peak}]{\includegraphics[width=0.4\textwidth]{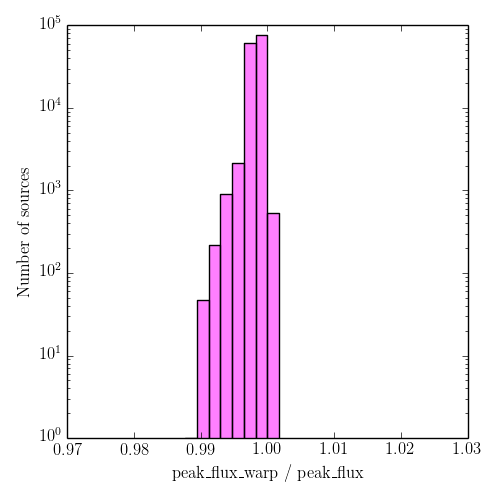}}\caption{Histograms of the measured source parameters in the ratio after:before warping.\label{fig:warp_sim_params_results}}
\end{figure*}

Histograms of the warped and input source parameters are shown in \fig~\ref{fig:warp_sim_params_results}. Only the integrated flux density shows any change greater than 0.5\%; peak flux density and source major and minor axes are essentially unchanged by the procedure. In contrast, the typical fitting uncertainty during source-finding on these sources is of order $\approx1.5$\,\%. Therefore, we can state that \fitswarp{} has a completely negligible effect on the source shapes.

To determine whether the change to integrated flux density is serious, i.e. worthy of some further post-facto correction, we examine its magnitude, and its correlation with the divergence of the vector field, since it is this "stretching" and the concomitant ``filling-in" by the interpolator that would change the integrated flux density of the sources.

\begin{figure}
\centering\includegraphics[width=\linewidth]{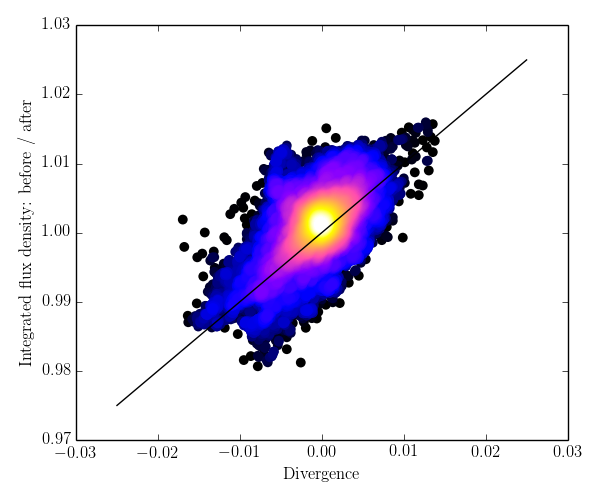}
\caption{The ratio of the integrated flux density of the sources before and after warping, $\frac{S'}{S}$, with respect to the divergence at the (unwarped) position of the sources. The colourscale shows the log of the number density of the points in arbitrary units, with the peak (white) corresponding to a density $10^4$ larger than the least dense areas (black). The black line shows a linear relationship between $\frac{S'}{S}$ and the divergence with unity offset (Equation~\ref{eq:int_vs_div}). Around this trend, a scatter of $\approx0.5$\% is observed.\label{fig:int_vs_div}}
\end{figure}

For every (central) source in every simulation, we measure the divergence at its (unwarped) location, and in \fig~\ref{fig:int_vs_div} we plot the ratio of warped and unwarped integrated flux densities against the divergence. As expected, there is a correlation: sources which have been ``stretched" where the field has positive divergence have slightly higher integrated flux densities, and sources which have been ``compressed" in regions with negative divergence have slightly lower integrated flux densities. A reasonable functional form for this is:
\begin{equation}
 \frac{S'}{S} = \dv{\Delta x}{x} + \dv{\Delta y}{y} + 1
 \label{eq:int_vs_div}
\end{equation}
as shown by the black line in \fig~\ref{fig:int_vs_div}.

While we have shown that there is a correlation, and therefore in principle a correction could be made, there are several arguments against doing so, at least in the MWA regime:
\begin{itemize}
\item{The warping effects applied here are real, typical of ionospheric conditions where $r_\mathrm{diff}>2.5$\,km above the MWA, and yet \fig~\ref{subfig:int} shows that only $\approx2$\% of sources have distortions $>1$\%: it would therefore be a relatively uncommon occurrence to need this correction;}
\item{The maximum magnitude of the change in integrated flux densities ($\pm1.5$\%) is well within the fitting and calibration errors to be expected in real radio data, and thus a correction would simply be ``in the noise'' (this is why we need to use simulated high S/N data to perform \ref{test2});}
\item{The implementation of such a correction would involve creating a map of the new point spread function induced by the warping; while such maps can be used by some software such as \textsc{Aegean}, it would be an extra complication for a typical user, for little gain;}
\item{Finally, for any given divergence value, there is a scatter of $\approx$0.5\% on the actual correction to the integrated flux density that would need to be applied; it is therefore a relatively low signal-to-noise correction, even for these high signal-to-noise sources.}
\end{itemize}

However, should the reader be operating outside of this regime, or otherwise decide that a correction would be useful, we provide an option to output the divergence maps, as well as maps of $\Delta x$ and $\Delta y$, which may be useful for debugging purposes.

\subsection{Fully simulated data}
\begin{figure}
\centering\includegraphics[width=0.95\linewidth]{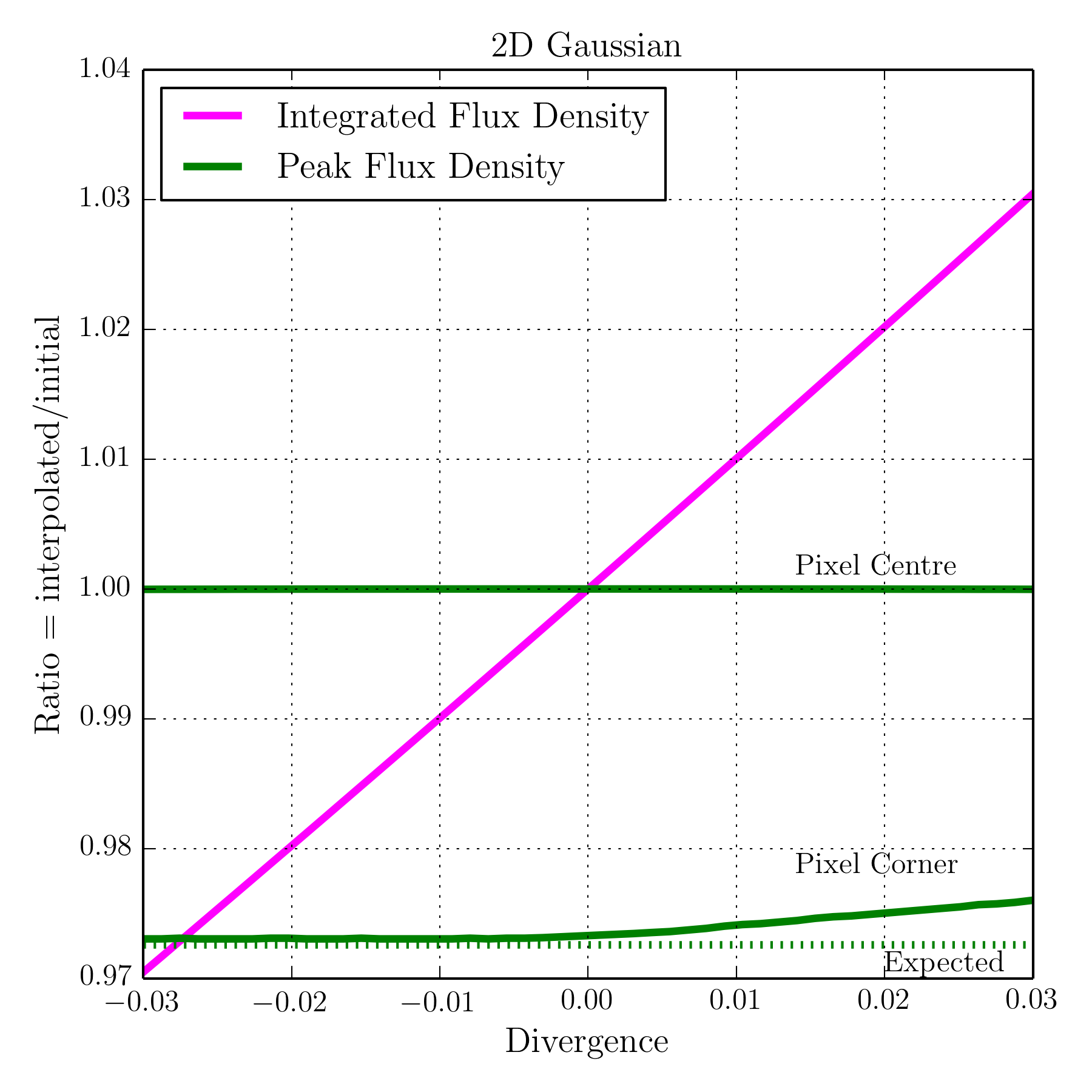}
\caption{The ratio of the integrated (summed) and peak (maximum) values for a 2D Gaussian which has been re-sampled and interpolated using scipy, over a range of divergence, and for two different source models.
The pixel center and pixel corner models produce the same results for the integrated flux density, but different results for the peak flux density.
The dashed line shows the expected peak flux density ratio for a source that falls on the corner of a pixel.
The range of divergence shown here is matched to that of \fig~\ref{fig:int_vs_div}.
Note that the same linear relationship is seen in both \fig~\ref{fig:int_vs_div} and this figure.
\label{fig:full_sim}}
\end{figure}

A final test was carried out in order to isolate any imaging or source characterization effects from the above work, and test only the effect of the image divergence; this final test is a simple simulation.
We start with a 2D Gaussian model with a FWHM of 5 pixels, which is a typical choice for radio imaging, and consistent with the data used in the previous tests.
Two models were included, a best case mode where the source is at the center of the pixel, and a worst case model where the source lands on the corner of a pixel.
The model data values are then shifted from their initial coordinates $(x,y)$ to a new set of coordinates $(x^\prime, y^\prime)$ using a mapping of:
\begin{equation}
(x,y) \mapsto (x^\prime, y^\prime) = ( x + (x-x_0)\times(1+\mathrm{div}/2) + \mathrm{eps}, y + (y-y_0)\times(1+\mathrm{div}/2)+ \mathrm{eps})
\end{equation}
where $(x_0,y_0)$ represent the central location of the Gaussian, div is the divergence of the map, and eps is an overall pixel offset to allow the center of the source to be offset from the center of the pixel.
An interpolator is initialized using {\sc scipy.interpolate.interp2d} with the initial data values and the remapped coordinates $(x^\prime, y^\prime)$.
The interpolator is then evaluated at the initial pixel coordinates $(x,y)$.
The model data and interpolated data are then summed to produce integrated pixel values, and the maximum of each is also calculated.
The ratio of remapped to model data for both the integrated and peak values are then recorded for a range of different divergences, and an eps of either 0 or 0.5.
\Fig\,\ref{fig:full_sim} shows the resulting curve of these ratios as a function of divergence.
The peak pixel value is expected to be unchanged when the model source is at the pixel center (eps=0).
Since the model source has a FWHM that is 10 pixels, a shift of 0.5 pixels in two directions represents a shift of 1/10th the FWHM ($=0.2355\sigma$), so the peak pixel flux is expected to drop to $\exp(-(0.2355)^2/2)$ when the model source is at the corner of the pixel (eps=0.5).
Both of these expectations are realized in our simulations.
Over the range of divergences measured in this work $(\pm 0.03)$ we predict a very nearly linear relation between the ratio of integrated values and divergence ($\mathrm{ratio} \simeq 1+\mathrm{div}$, c.f \eqn\,\ref{eq:int_vs_div}), whereas the ratio of peak values is constant at unity.
When \Fig\,\ref{fig:full_sim} is expanded to a much larger range of divergence it is clear that the relationship between the integrated ratio and divergence is quadratic.

The results of \Fig\,\ref{fig:full_sim} are clear: we do not expect the peak flux density to be significantly affected by \fitswarp{}, and the integrated flux densities will differ by an amount that is essentially linearly proportional to the divergence.
The agreement between the measurements of \Fig\,\ref{fig:int_vs_div} and the expectations in \Fig\,\ref{fig:full_sim} demonstrates the general applicability of the results in the previous two sections.

\section{Conclusions}\label{sec:conclusions}

We have demonstrated an algorithm which can de-distort the refractive effects of the ionosphere on astronomical FITS images, given a reliable input reference catalogue. It is useful in ionospheric regimes where the scale size of coherent ionospheric features is larger than that of the longest baseline of the telescope being used, i.e. $r_\mathrm{diff}>D$.


We note that this algorithm is general-purpose to de-distort any image distorted by some vector field which is sampled by some sparse pierce-points. This may have applications outside the field of low-frequency radio astronomy. The authors welcome contact and discussions on making the code more general-purpose.

\appendix

\begin{table}
\caption{MWA observations used in this work.}
\label{tab:obsids}
\centering
\begin{tabular}{cccc}
1061673800 & 1062276944 & 1062880096 & 1063483240 \\
1064689544 & 1067102136 & 1068480760 & 1068911584 \\
1069514728 & 1070117880 & 1091400432 & 1092520568 \\
1093037552 & 1094330008 & 1094760832 & 1095536312 \\
1096139456 & 1097345752 & 1098207400 & 1099069040 \\
\end{tabular}
\end{table}

\section{Implementation details}

\subsection{Memory considerations}

Python allows relatively little control over memory use, yet the \fitswarp{} algorithm has been and will continue to be used on supercomputers, which typically have no swap memory, and kill processes which require more memory than present on the node. There are two memory-hogging steps in the current python implementation which act as a bottleneck for ``reasonable'' systems (e.g. typical current supercomputer nodes, desktop and laptop computers):
\begin{itemize}
\item{The calculation of the position shifts $\Delta x$,$\Delta y$ for $\approx$millions of pixel positions (\ref{callmodel} above);}
\item{The interpolation to generate the new $F'$ (\ref{interpolate} above).}
\end{itemize}

In order to constrain the processing within the available memory of the system on which the algorithm runs, we use \textsc{psutil}\footnote{https://psutil.readthedocs.io/en/latest/} to determine the available system memory, multiply it by a padding factor of 0.75 to account for potential overestimation at the moment of measurement, and divide the processing of \ref{callmodel} and \ref{interpolate} into multiple ``strides'', the number depending how many will fit into the available memory. Since these stages are memory-limited, they operate on a single CPU thread. This accounts for over 90\% of the processing time. Therefore the speed of the program can be dramatically improved by running it on larger-memory systems, or by some future improvement that reduces the amount of memory needed.

\subsection{Interpolator}

We chose the Clough-Tocher 2D interpolator from the \textsc{SciPy} package to interpolate the images. It is useful because it does not require the inputs to be on a regular grid, and produces extremely smooth results, which minimize changes to the images. However, the \textsc{SciPy} implementation does have at least one issue: if there are any NaNs in the image, the interpolation will produce large areas of NaNs in the resulting image, in seemingly random areas. However in radio astronomy, it is common to have some areas which are NaN, for instance if they lie out of the coverage of the image projection, and thus have undefined WCS co-ordinates (even if they have perfectly valid pixel co-ordinates).

To work around this, just before \ref{interpolate}, we select all pixels in the image which are value NaN and set their values to zero. We re-set these zeroed regions to NaN after this same step has finished. In addition, we set to NaN a small (default of 10~pixels) exclusion zone around the edges of the images to avoid outputting artefacts due to incomplete interpolation.




\section*{Acknowledgements}
We acknowledge the work and support of the developers of the following following python packages: Astropy \citet{TheAstropyCollaboration2013,2018arXiv180102634T}, Numpy \citep{vaderwalt_numpy_2011} and Scipy \citep{Jones_scipy_2001}.
To perform catalogue and image analysis, this work made use of DS9\footnote{\texttt{http://ds9.si.edu/site/Home.html}} and Topcat \citep{Taylor_topcat_2005}.



\newcommand{\pasa}{PASA}
\newcommand{\aj}{AJ}
\newcommand{\apj}{ApJ}
\newcommand{\apjs}{ApJS}
\newcommand{\apjl}{ApJL}
\newcommand{\aap}{A{\&}A}
\newcommand{\aaps}{A{\&}AS}
\newcommand{\mnras}{MNRAS}
\newcommand{\araa}{ARAA}
\newcommand{\pasp}{PASP}
\newcommand{\nat}{Nature}
\newcommand{\grl}{Geophysical Research Letters}

\bibliographystyle{model2-names}
\bibliography{fitswarp.bib}

\begin{thebibliography}{28}
\expandafter\ifx\csname natexlab\endcsname\relax\def\natexlab#1{#1}\fi
\expandafter\ifx\csname url\endcsname\relax
  \def\url#1{\texttt{#1}}\fi
\expandafter\ifx\csname urlprefix\endcsname\relax\def\urlprefix{URL }\fi
\providecommand{\eprint}[2][]{\url{#2}}
\providecommand{\bibinfo}[2]{#2}
\ifx\xfnm\relax \def\xfnm[#1]{\unskip,\space#1}\fi
\bibitem[{{Bertin}(2006)}]{2006ASPC..351..112B}
\bibinfo{author}{{Bertin}, E.}, \bibinfo{year}{2006}.
\newblock \bibinfo{title}{{Automatic Astrometric and Photometric Calibration
  with SCAMP}}, in: \bibinfo{editor}{{Gabriel}, C.},
  \bibinfo{editor}{{Arviset}, C.}, \bibinfo{editor}{{Ponz}, D.},
  \bibinfo{editor}{{Enrique}, S.} (Eds.), \bibinfo{booktitle}{Astronomical Data
  Analysis Software and Systems XV}, p. \bibinfo{pages}{112}.
\bibitem[{{Cotton}(2005)}]{2005ASPC..345..337C}
\bibinfo{author}{{Cotton}, W.D.}, \bibinfo{year}{2005}.
\newblock \bibinfo{title}{{Lessons from the VLA Long Wavelength Sky Survey
  (VLSS)}}, in: \bibinfo{editor}{{Kassim}, N.}, \bibinfo{editor}{{Perez}, M.},
  \bibinfo{editor}{{Junor}, W.}, \bibinfo{editor}{{Henning}, P.} (Eds.),
  \bibinfo{booktitle}{From Clark Lake to the Long Wavelength Array: Bill
  Erickson's Radio Science}, p. \bibinfo{pages}{337}.
\bibitem[{{Davies} and {Kasper}(2012)}]{2012ARA&A..50..305D}
\bibinfo{author}{{Davies}, R.}, \bibinfo{author}{{Kasper}, M.},
  \bibinfo{year}{2012}.
\newblock \bibinfo{title}{{Adaptive Optics for Astronomy}}.
\newblock \bibinfo{journal}{\araa} \bibinfo{volume}{50},
  \bibinfo{pages}{305--351}.
\newblock \eprint{1201.5741}.
\bibitem[{{Fomalont}(1999)}]{1999ASPC..180..301F}
\bibinfo{author}{{Fomalont}, E.B.}, \bibinfo{year}{1999}.
\newblock \bibinfo{title}{{Image Analysis}}, in: \bibinfo{editor}{{Taylor},
  G.B.}, \bibinfo{editor}{{Carilli}, C.L.}, \bibinfo{editor}{{Perley}, R.A.}
  (Eds.), \bibinfo{booktitle}{Synthesis Imaging in Radio Astronomy II}, p.
  \bibinfo{pages}{301}.
\bibitem[{Hancock et~al.(2012)Hancock, Murphy, Gaensler, Hopkins and
  Curran}]{Hancock_compact_2012}
\bibinfo{author}{Hancock, P.}, \bibinfo{author}{Murphy, T.},
  \bibinfo{author}{Gaensler, B.}, \bibinfo{author}{Hopkins, A.},
  \bibinfo{author}{Curran, J.}, \bibinfo{year}{2012}.
\newblock \bibinfo{title}{{Compact continuum source finding for next generation
  radio surveys}}.
\newblock \bibinfo{journal}{Monthly Notices of the Royal Astronomical Society}
  \bibinfo{volume}{422}.
\bibitem[{{Hancock} et~al.(2018){Hancock}, {Trott} and
  {Hurley-Walker}}]{2018PASA...35...11H}
\bibinfo{author}{{Hancock}, P.J.}, \bibinfo{author}{{Trott}, C.M.},
  \bibinfo{author}{{Hurley-Walker}, N.}, \bibinfo{year}{2018}.
\newblock \bibinfo{title}{{Source Finding in the Era of the SKA (Precursors):
  Aegean 2.0}}.
\newblock \bibinfo{journal}{\pasa} \bibinfo{volume}{35}, \bibinfo{pages}{e011}.
\newblock \eprint{1801.05548}.
\bibitem[{{Hurley-Walker} et~al.(2017){Hurley-Walker}, {Callingham}, {Hancock},
  {Franzen}, {Hindson} et~al.}]{GLEAMEGC}
\bibinfo{author}{{Hurley-Walker}, N.}, \bibinfo{author}{{Callingham}, J.R.},
  \bibinfo{author}{{Hancock}, P.J.}, \bibinfo{author}{{Franzen}, T.M.O.},
  \bibinfo{author}{{Hindson}, L.}, et~al., \bibinfo{year}{2017}.
\newblock \bibinfo{title}{{GaLactic and Extragalactic All-sky Murchison
  Widefield Array (GLEAM) survey - I. A low-frequency extragalactic
  catalogue}}.
\newblock \bibinfo{journal}{MNRAS} \bibinfo{volume}{464},
  \bibinfo{pages}{1146--1167}.
\newblock \eprint{1610.08318}.
\bibitem[{{Intema} et~al.(2009){Intema}, {van der Tol}, {Cotton}, {Cohen}, {van
  Bemmel} and {R{\"o}ttgering}}]{2009A&A...501.1185I}
\bibinfo{author}{{Intema}, H.T.}, \bibinfo{author}{{van der Tol}, S.},
  \bibinfo{author}{{Cotton}, W.D.}, \bibinfo{author}{{Cohen}, A.S.},
  \bibinfo{author}{{van Bemmel}, I.M.}, \bibinfo{author}{{R{\"o}ttgering},
  H.J.A.}, \bibinfo{year}{2009}.
\newblock \bibinfo{title}{{Ionospheric calibration of low frequency radio
  interferometric observations using the peeling scheme. I. Method description
  and first results}}.
\newblock \bibinfo{journal}{\aap} \bibinfo{volume}{501},
  \bibinfo{pages}{1185--1205}.
\newblock \eprint{0904.3975}.
\bibitem[{Jones et~al.(2001--)Jones, Oliphant, Peterson
  et~al.}]{Jones_scipy_2001}
\bibinfo{author}{Jones, E.}, \bibinfo{author}{Oliphant, T.},
  \bibinfo{author}{Peterson, P.}, et~al., \bibinfo{year}{2001--}.
\newblock \bibinfo{title}{{SciPy}: Open source scientific tools for {Python}}.
\newblock \bibinfo{note}{[Online; accessed 2017-08-04]}.
\bibitem[{{Jordan} et~al.(2017){Jordan}, {Murray}, {Trott}, {Wayth},
  {Mitchell}, {Rahimi}, {Pindor}, {Procopio} and
  {Morgan}}]{2017MNRAS.471.3974J}
\bibinfo{author}{{Jordan}, C.H.}, \bibinfo{author}{{Murray}, S.},
  \bibinfo{author}{{Trott}, C.M.}, \bibinfo{author}{{Wayth}, R.B.},
  \bibinfo{author}{{Mitchell}, D.A.}, \bibinfo{author}{{Rahimi}, M.},
  \bibinfo{author}{{Pindor}, B.}, \bibinfo{author}{{Procopio}, P.},
  \bibinfo{author}{{Morgan}, J.}, \bibinfo{year}{2017}.
\newblock \bibinfo{title}{{Characterization of the ionosphere above the
  Murchison Radio Observatory using the Murchison Widefield Array}}.
\newblock \bibinfo{journal}{\mnras} \bibinfo{volume}{471},
  \bibinfo{pages}{3974--3987}.
\newblock \eprint{1707.04978}.
\bibitem[{{Loi} et~al.(2015){Loi}, {Murphy}, {Cairns}, {Menk}, {Waters},
  {Erickson}, {Trott}, {Hurley-Walker}, {Morgan}, {Lenc}, {Offringa}, {Bell},
  {Ekers}, {Gaensler}, {Lonsdale}, {Feng}, {Hancock}, {Kaplan}, {Bernardi},
  {Bowman}, {Briggs}, {Cappallo}, {Deshpande}, {Greenhill}, {Hazelton},
  {Johnston-Hollitt}, {McWhirter}, {Mitchell}, {Morales}, {Morgan}, {Oberoi},
  {Ord}, {Prabu}, {Shankar}, {Srivani}, {Subrahmanyan}, {Tingay}, {Wayth},
  {Webster}, {Williams} and {Williams}}]{2015GeoRL..42.3707L}
\bibinfo{author}{{Loi}, S.T.}, \bibinfo{author}{{Murphy}, T.},
  \bibinfo{author}{{Cairns}, I.H.}, \bibinfo{author}{{Menk}, F.W.},
  \bibinfo{author}{{Waters}, C.L.}, \bibinfo{author}{{Erickson}, P.J.},
  \bibinfo{author}{{Trott}, C.M.}, \bibinfo{author}{{Hurley-Walker}, N.},
  \bibinfo{author}{{Morgan}, J.}, \bibinfo{author}{{Lenc}, E.},
  \bibinfo{author}{{Offringa}, A.R.}, \bibinfo{author}{{Bell}, M.E.},
  \bibinfo{author}{{Ekers}, R.D.}, \bibinfo{author}{{Gaensler}, B.M.},
  \bibinfo{author}{{Lonsdale}, C.J.}, \bibinfo{author}{{Feng}, L.},
  \bibinfo{author}{{Hancock}, P.J.}, \bibinfo{author}{{Kaplan}, D.L.},
  \bibinfo{author}{{Bernardi}, G.}, \bibinfo{author}{{Bowman}, J.D.},
  \bibinfo{author}{{Briggs}, F.}, \bibinfo{author}{{Cappallo}, R.J.},
  \bibinfo{author}{{Deshpande}, A.A.}, \bibinfo{author}{{Greenhill}, L.J.},
  \bibinfo{author}{{Hazelton}, B.J.}, \bibinfo{author}{{Johnston-Hollitt}, M.},
  \bibinfo{author}{{McWhirter}, S.R.}, \bibinfo{author}{{Mitchell}, D.A.},
  \bibinfo{author}{{Morales}, M.F.}, \bibinfo{author}{{Morgan}, E.},
  \bibinfo{author}{{Oberoi}, D.}, \bibinfo{author}{{Ord}, S.M.},
  \bibinfo{author}{{Prabu}, T.}, \bibinfo{author}{{Shankar}, N.U.},
  \bibinfo{author}{{Srivani}, K.S.}, \bibinfo{author}{{Subrahmanyan}, R.},
  \bibinfo{author}{{Tingay}, S.J.}, \bibinfo{author}{{Wayth}, R.B.},
  \bibinfo{author}{{Webster}, R.L.}, \bibinfo{author}{{Williams}, A.},
  \bibinfo{author}{{Williams}, C.L.}, \bibinfo{year}{2015}.
\newblock \bibinfo{title}{{Real-time imaging of density ducts between the
  plasmasphere and ionosphere}}.
\newblock \bibinfo{journal}{\grl} \bibinfo{volume}{42},
  \bibinfo{pages}{3707--3714}.
\newblock \eprint{1504.06470}.
\bibitem[{{Lonsdale}(2005)}]{2005ASPC..345..399L}
\bibinfo{author}{{Lonsdale}, C.J.}, \bibinfo{year}{2005}.
\newblock \bibinfo{title}{{Configuration Considerations for Low Frequency
  Arrays}}, in: \bibinfo{editor}{{Kassim}, N.}, \bibinfo{editor}{{Perez}, M.},
  \bibinfo{editor}{{Junor}, W.}, \bibinfo{editor}{{Henning}, P.} (Eds.),
  \bibinfo{booktitle}{From Clark Lake to the Long Wavelength Array: Bill
  Erickson's Radio Science}, p. \bibinfo{pages}{399}.
\bibitem[{{Lonsdale} et~al.(2009){Lonsdale}, {Cappallo}, {Morales}
  et~al.}]{Lonsdale}
\bibinfo{author}{{Lonsdale}, C.J.}, \bibinfo{author}{{Cappallo}, R.J.},
  \bibinfo{author}{{Morales}, M.F.}, et~al., \bibinfo{year}{2009}.
\newblock \bibinfo{title}{{The Murchison Widefield Array: Design Overview}}.
\newblock \bibinfo{journal}{IEEE Proceedings} \bibinfo{volume}{97},
  \bibinfo{pages}{1497--1506}.
\newblock \eprint{0903.1828}.
\bibitem[{{Mevius} et~al.(2016){Mevius}, {van der Tol}, {Pandey}, {Vedantham},
  {Brentjens}, {de Bruyn}, {Abdalla}, {Asad}, {Bregman}, {Brouw}, {Bus},
  {Chapman}, {Ciardi}, {Fernandez}, {Ghosh}, {Harker}, {Iliev}, {Jeli{\'c}},
  {Kazemi}, {Koopmans}, {Noordam}, {Offringa}, {Patil}, {van Weeren},
  {Wijnholds}, {Yatawatta} and {Zaroubi}}]{2016RaSc...51..927M}
\bibinfo{author}{{Mevius}, M.}, \bibinfo{author}{{van der Tol}, S.},
  \bibinfo{author}{{Pandey}, V.N.}, \bibinfo{author}{{Vedantham}, H.K.},
  \bibinfo{author}{{Brentjens}, M.A.}, \bibinfo{author}{{de Bruyn}, A.G.},
  \bibinfo{author}{{Abdalla}, F.B.}, \bibinfo{author}{{Asad}, K.M.B.},
  \bibinfo{author}{{Bregman}, J.D.}, \bibinfo{author}{{Brouw}, W.N.},
  \bibinfo{author}{{Bus}, S.}, \bibinfo{author}{{Chapman}, E.},
  \bibinfo{author}{{Ciardi}, B.}, \bibinfo{author}{{Fernandez}, E.R.},
  \bibinfo{author}{{Ghosh}, A.}, \bibinfo{author}{{Harker}, G.},
  \bibinfo{author}{{Iliev}, I.T.}, \bibinfo{author}{{Jeli{\'c}}, V.},
  \bibinfo{author}{{Kazemi}, S.}, \bibinfo{author}{{Koopmans}, L.V.E.},
  \bibinfo{author}{{Noordam}, J.E.}, \bibinfo{author}{{Offringa}, A.R.},
  \bibinfo{author}{{Patil}, A.H.}, \bibinfo{author}{{van Weeren}, R.J.},
  \bibinfo{author}{{Wijnholds}, S.}, \bibinfo{author}{{Yatawatta}, S.},
  \bibinfo{author}{{Zaroubi}, S.}, \bibinfo{year}{2016}.
\newblock \bibinfo{title}{{Probing ionospheric structures using the LOFAR radio
  telescope}}.
\newblock \bibinfo{journal}{Radio Science} \bibinfo{volume}{51},
  \bibinfo{pages}{927--941}.
\newblock \eprint{1606.04683}.
\bibitem[{{Offringa} et~al.(2014){Offringa}, {McKinley}, {Hurley-Walker},
  {Briggs}, {Wayth}, {Kaplan}, {Bell}, {Feng}, {Neben}, {Hughes}, {Rhee},
  {Murphy}, {Bhat}, {Bernardi}, {Bowman}, {Cappallo}, {Corey}, {Deshpande},
  {Emrich}, {Ewall-Wice}, {Gaensler}, {Goeke}, {Greenhill}, {Hazelton},
  {Hindson}, {Johnston-Hollitt}, {Jacobs}, {Kasper}, {Kratzenberg}, {Lenc},
  {Lonsdale}, {Lynch}, {McWhirter}, {Mitchell}, {Morales}, {Morgan},
  {Kudryavtseva}, {Oberoi}, {Ord}, {Pindor}, {Procopio}, {Prabu}, {Riding},
  {Roshi}, {Shankar}, {Srivani}, {Subrahmanyan}, {Tingay}, {Waterson},
  {Webster}, {Whitney}, {Williams} and {Williams}}]{2014MNRAS.444..606O}
\bibinfo{author}{{Offringa}, A.R.}, \bibinfo{author}{{McKinley}, B.},
  \bibinfo{author}{{Hurley-Walker}, N.}, \bibinfo{author}{{Briggs}, F.H.},
  \bibinfo{author}{{Wayth}, R.B.}, \bibinfo{author}{{Kaplan}, D.L.},
  \bibinfo{author}{{Bell}, M.E.}, \bibinfo{author}{{Feng}, L.},
  \bibinfo{author}{{Neben}, A.R.}, \bibinfo{author}{{Hughes}, J.D.},
  \bibinfo{author}{{Rhee}, J.}, \bibinfo{author}{{Murphy}, T.},
  \bibinfo{author}{{Bhat}, N.D.R.}, \bibinfo{author}{{Bernardi}, G.},
  \bibinfo{author}{{Bowman}, J.D.}, \bibinfo{author}{{Cappallo}, R.J.},
  \bibinfo{author}{{Corey}, B.E.}, \bibinfo{author}{{Deshpande}, A.A.},
  \bibinfo{author}{{Emrich}, D.}, \bibinfo{author}{{Ewall-Wice}, A.},
  \bibinfo{author}{{Gaensler}, B.M.}, \bibinfo{author}{{Goeke}, R.},
  \bibinfo{author}{{Greenhill}, L.J.}, \bibinfo{author}{{Hazelton}, B.J.},
  \bibinfo{author}{{Hindson}, L.}, \bibinfo{author}{{Johnston-Hollitt}, M.},
  \bibinfo{author}{{Jacobs}, D.C.}, \bibinfo{author}{{Kasper}, J.C.},
  \bibinfo{author}{{Kratzenberg}, E.}, \bibinfo{author}{{Lenc}, E.},
  \bibinfo{author}{{Lonsdale}, C.J.}, \bibinfo{author}{{Lynch}, M.J.},
  \bibinfo{author}{{McWhirter}, S.R.}, \bibinfo{author}{{Mitchell}, D.A.},
  \bibinfo{author}{{Morales}, M.F.}, \bibinfo{author}{{Morgan}, E.},
  \bibinfo{author}{{Kudryavtseva}, N.}, \bibinfo{author}{{Oberoi}, D.},
  \bibinfo{author}{{Ord}, S.M.}, \bibinfo{author}{{Pindor}, B.},
  \bibinfo{author}{{Procopio}, P.}, \bibinfo{author}{{Prabu}, T.},
  \bibinfo{author}{{Riding}, J.}, \bibinfo{author}{{Roshi}, D.A.},
  \bibinfo{author}{{Shankar}, N.U.}, \bibinfo{author}{{Srivani}, K.S.},
  \bibinfo{author}{{Subrahmanyan}, R.}, \bibinfo{author}{{Tingay}, S.J.},
  \bibinfo{author}{{Waterson}, M.}, \bibinfo{author}{{Webster}, R.L.},
  \bibinfo{author}{{Whitney}, A.R.}, \bibinfo{author}{{Williams}, A.},
  \bibinfo{author}{{Williams}, C.L.}, \bibinfo{year}{2014}.
\newblock \bibinfo{title}{{WSCLEAN: an implementation of a fast, generic
  wide-field imager for radio astronomy}}.
\newblock \bibinfo{journal}{\mnras} \bibinfo{volume}{444},
  \bibinfo{pages}{606--619}.
\newblock \eprint{1407.1943}.
\bibitem[{{Offringa} et~al.(2016){Offringa}, {Trott}, {Hurley-Walker},
  {Johnston-Hollitt}, {McKinley}, {Barry}, {Beardsley}, {Bowman}, {Briggs},
  {Carroll}, {Dillon}, {Ewall-Wice}, {Feng}, {Gaensler}, {Greenhill},
  {Hazelton}, {Hewitt}, {Jacobs}, {Kim}, {Kittiwisit}, {Lenc}, {Line}, {Loeb},
  {Mitchell}, {Morales}, {Neben}, {Paul}, {Pindor}, {Pober}, {Procopio},
  {Riding}, {Sethi}, {Shankar}, {Subrahmanyan}, {Sullivan}, {Tegmark},
  {Thyagarajan}, {Tingay}, {Wayth}, {Webster} and
  {Wyithe}}]{2016MNRAS.458.1057O}
\bibinfo{author}{{Offringa}, A.R.}, \bibinfo{author}{{Trott}, C.M.},
  \bibinfo{author}{{Hurley-Walker}, N.}, \bibinfo{author}{{Johnston-Hollitt},
  M.}, \bibinfo{author}{{McKinley}, B.}, \bibinfo{author}{{Barry}, N.},
  \bibinfo{author}{{Beardsley}, A.P.}, \bibinfo{author}{{Bowman}, J.D.},
  \bibinfo{author}{{Briggs}, F.}, \bibinfo{author}{{Carroll}, P.},
  \bibinfo{author}{{Dillon}, J.S.}, \bibinfo{author}{{Ewall-Wice}, A.},
  \bibinfo{author}{{Feng}, L.}, \bibinfo{author}{{Gaensler}, B.M.},
  \bibinfo{author}{{Greenhill}, L.J.}, \bibinfo{author}{{Hazelton}, B.J.},
  \bibinfo{author}{{Hewitt}, J.N.}, \bibinfo{author}{{Jacobs}, D.C.},
  \bibinfo{author}{{Kim}, H.S.}, \bibinfo{author}{{Kittiwisit}, P.},
  \bibinfo{author}{{Lenc}, E.}, \bibinfo{author}{{Line}, J.},
  \bibinfo{author}{{Loeb}, A.}, \bibinfo{author}{{Mitchell}, D.A.},
  \bibinfo{author}{{Morales}, M.F.}, \bibinfo{author}{{Neben}, A.R.},
  \bibinfo{author}{{Paul}, S.}, \bibinfo{author}{{Pindor}, B.},
  \bibinfo{author}{{Pober}, J.C.}, \bibinfo{author}{{Procopio}, P.},
  \bibinfo{author}{{Riding}, J.}, \bibinfo{author}{{Sethi}, S.K.},
  \bibinfo{author}{{Shankar}, N.U.}, \bibinfo{author}{{Subrahmanyan}, R.},
  \bibinfo{author}{{Sullivan}, I.S.}, \bibinfo{author}{{Tegmark}, M.},
  \bibinfo{author}{{Thyagarajan}, N.}, \bibinfo{author}{{Tingay}, S.J.},
  \bibinfo{author}{{Wayth}, R.B.}, \bibinfo{author}{{Webster}, R.L.},
  \bibinfo{author}{{Wyithe}, J.S.B.}, \bibinfo{year}{2016}.
\newblock \bibinfo{title}{{Parametrizing Epoch of Reionization foregrounds: a
  deep survey of low-frequency point-source spectra with the Murchison
  Widefield Array}}.
\newblock \bibinfo{journal}{\mnras} \bibinfo{volume}{458},
  \bibinfo{pages}{1057--1070}.
\newblock \eprint{1602.02247}.
\bibitem[{{Swarup} et~al.(1991){Swarup}, {Ananthakrishnan}, {Kapahi}, {Rao},
  {Subrahmanya} and {Kulkarni}}]{1991CuSc...60...95S}
\bibinfo{author}{{Swarup}, G.}, \bibinfo{author}{{Ananthakrishnan}, S.},
  \bibinfo{author}{{Kapahi}, V.K.}, \bibinfo{author}{{Rao}, A.P.},
  \bibinfo{author}{{Subrahmanya}, C.R.}, \bibinfo{author}{{Kulkarni}, V.K.},
  \bibinfo{year}{1991}.
\newblock \bibinfo{title}{{The Giant Metre-Wave Radio Telescope}}.
\newblock \bibinfo{journal}{Current Science, Vol.~60, NO.2/JAN25, P.~95, 1991}
  \bibinfo{volume}{60}, \bibinfo{pages}{95}.
\bibitem[{{Taylor} et~al.(2012){Taylor}, {Ellingson}, {Kassim}, {Craig},
  {Dowell}, {Wolfe}, {Hartman}, {Bernardi}, {Clarke}, {Cohen} and {et
  al}}]{Taylor}
\bibinfo{author}{{Taylor}, G.B.}, \bibinfo{author}{{Ellingson}, S.W.},
  \bibinfo{author}{{Kassim}, N.E.}, \bibinfo{author}{{Craig}, J.},
  \bibinfo{author}{{Dowell}, J.}, \bibinfo{author}{{Wolfe}, C.N.},
  \bibinfo{author}{{Hartman}, J.}, \bibinfo{author}{{Bernardi}, G.},
  \bibinfo{author}{{Clarke}, T.}, \bibinfo{author}{{Cohen}, A.},
  \bibinfo{author}{{et al}}, \bibinfo{year}{2012}.
\newblock \bibinfo{title}{{First Light for the First Station of the Long
  Wavelength Array}}.
\newblock \bibinfo{journal}{JAI} \bibinfo{volume}{1}, \bibinfo{pages}{50004}.
\newblock \eprint{1206.6733}.
\bibitem[{{Taylor}(2005)}]{Taylor_topcat_2005}
\bibinfo{author}{{Taylor}, M.B.}, \bibinfo{year}{2005}.
\newblock \bibinfo{title}{{TOPCAT and STIL: Starlink Table/VOTable Processing
  Software}}, in: \bibinfo{editor}{{Shopbell}, P.}, \bibinfo{editor}{{Britton},
  M.}, \bibinfo{editor}{{Ebert}, R.} (Eds.), \bibinfo{booktitle}{Astronomical
  Data Analysis Software and Systems XIV}, p.~\bibinfo{pages}{29}.
\bibitem[{{The Astropy Collaboration} et~al.(2018){The Astropy Collaboration},
  {Price-Whelan}, {Sip{\H o}cz}, {G{\"u}nther}, {Lim}, {Crawford}, {Conseil},
  {Shupe}, {Craig}, {Dencheva}, {Ginsburg}, {VanderPlas}, {Bradley},
  {P{\'e}rez-Su{\'a}rez}, {de Val-Borro}, {Aldcroft}, {Cruz}, {Robitaille},
  {Tollerud}, {Ardelean}, {Babej}, {Bachetti}, {Bakanov}, {Bamford},
  {Barentsen}, {Barmby}, {Baumbach}, {Berry}, {Biscani}, {Boquien}, {Bostroem},
  {Bouma}, {Brammer}, {Bray}, {Breytenbach}, {Buddelmeijer}, {Burke},
  {Calderone}, {Cano Rodr{\'{\i}}guez}, {Cara}, {Cardoso}, {Cheedella},
  {Copin}, {Crichton}, {D{\'A}vella}, {Deil}, {Depagne}, {Dietrich}, {Donath},
  {Droettboom}, {Earl}, {Erben}, {Fabbro}, {Ferreira}, {Finethy}, {Fox},
  {Garrison}, {Gibbons}, {Goldstein}, {Gommers}, {Greco}, {Greenfield},
  {Groener}, {Grollier}, {Hagen}, {Hirst}, {Homeier}, {Horton}, {Hosseinzadeh},
  {Hu}, {Hunkeler}, {Ivezi{\'c}}, {Jain}, {Jenness}, {Kanarek}, {Kendrew},
  {Kern}, {Kerzendorf}, {Khvalko}, {King}, {Kirkby}, {Kulkarni}, {Kumar},
  {Lee}, {Lenz}, {Littlefair}, {Ma}, {Macleod}, {Mastropietro}, {McCully},
  {Montagnac}, {Morris}, {Mueller}, {Mumford}, {Muna}, {Murphy}, {Nelson},
  {Nguyen}, {Ninan}, {N{\"o}the}, {Ogaz}, {Oh}, {Parejko}, {Parley}, {Pascual},
  {Patil}, {Patil}, {Plunkett}, {Prochaska}, {Rastogi}, {Reddy Janga},
  {Sabater}, {Sakurikar}, {Seifert}, {Sherbert}, {Sherwood-Taylor}, {Shih},
  {Sick}, {Silbiger}, {Singanamalla}, {Singer}, {Sladen}, {Sooley},
  {Sornarajah}, {Streicher}, {Teuben}, {Thomas}, {Tremblay}, {Turner},
  {Terr{\'o}n}, {van Kerkwijk}, {de la Vega}, {Watkins}, {Weaver}, {Whitmore},
  {Woillez} and {Zabalza}}]{2018arXiv180102634T}
\bibinfo{author}{{The Astropy Collaboration}}, \bibinfo{author}{{Price-Whelan},
  A.M.}, \bibinfo{author}{{Sip{\H o}cz}, B.M.}, \bibinfo{author}{{G{\"u}nther},
  H.M.}, \bibinfo{author}{{Lim}, P.L.}, \bibinfo{author}{{Crawford}, S.M.},
  \bibinfo{author}{{Conseil}, S.}, \bibinfo{author}{{Shupe}, D.L.},
  \bibinfo{author}{{Craig}, M.W.}, \bibinfo{author}{{Dencheva}, N.},
  \bibinfo{author}{{Ginsburg}, A.}, \bibinfo{author}{{VanderPlas}, J.T.},
  \bibinfo{author}{{Bradley}, L.D.}, \bibinfo{author}{{P{\'e}rez-Su{\'a}rez},
  D.}, \bibinfo{author}{{de Val-Borro}, M.}, \bibinfo{author}{{Aldcroft},
  T.L.}, \bibinfo{author}{{Cruz}, K.L.}, \bibinfo{author}{{Robitaille}, T.P.},
  \bibinfo{author}{{Tollerud}, E.J.}, \bibinfo{author}{{Ardelean}, C.},
  \bibinfo{author}{{Babej}, T.}, \bibinfo{author}{{Bachetti}, M.},
  \bibinfo{author}{{Bakanov}, A.V.}, \bibinfo{author}{{Bamford}, S.P.},
  \bibinfo{author}{{Barentsen}, G.}, \bibinfo{author}{{Barmby}, P.},
  \bibinfo{author}{{Baumbach}, A.}, \bibinfo{author}{{Berry}, K.L.},
  \bibinfo{author}{{Biscani}, F.}, \bibinfo{author}{{Boquien}, M.},
  \bibinfo{author}{{Bostroem}, K.A.}, \bibinfo{author}{{Bouma}, L.G.},
  \bibinfo{author}{{Brammer}, G.B.}, \bibinfo{author}{{Bray}, E.M.},
  \bibinfo{author}{{Breytenbach}, H.}, \bibinfo{author}{{Buddelmeijer}, H.},
  \bibinfo{author}{{Burke}, D.J.}, \bibinfo{author}{{Calderone}, G.},
  \bibinfo{author}{{Cano Rodr{\'{\i}}guez}, J.L.}, \bibinfo{author}{{Cara},
  M.}, \bibinfo{author}{{Cardoso}, J.V.M.}, \bibinfo{author}{{Cheedella}, S.},
  \bibinfo{author}{{Copin}, Y.}, \bibinfo{author}{{Crichton}, D.},
  \bibinfo{author}{{D{\'A}vella}, D.}, \bibinfo{author}{{Deil}, C.},
  \bibinfo{author}{{Depagne}, {\'E}.}, \bibinfo{author}{{Dietrich}, J.P.},
  \bibinfo{author}{{Donath}, A.}, \bibinfo{author}{{Droettboom}, M.},
  \bibinfo{author}{{Earl}, N.}, \bibinfo{author}{{Erben}, T.},
  \bibinfo{author}{{Fabbro}, S.}, \bibinfo{author}{{Ferreira}, L.A.},
  \bibinfo{author}{{Finethy}, T.}, \bibinfo{author}{{Fox}, R.T.},
  \bibinfo{author}{{Garrison}, L.H.}, \bibinfo{author}{{Gibbons}, S.L.J.},
  \bibinfo{author}{{Goldstein}, D.A.}, \bibinfo{author}{{Gommers}, R.},
  \bibinfo{author}{{Greco}, J.P.}, \bibinfo{author}{{Greenfield}, P.},
  \bibinfo{author}{{Groener}, A.M.}, \bibinfo{author}{{Grollier}, F.},
  \bibinfo{author}{{Hagen}, A.}, \bibinfo{author}{{Hirst}, P.},
  \bibinfo{author}{{Homeier}, D.}, \bibinfo{author}{{Horton}, A.J.},
  \bibinfo{author}{{Hosseinzadeh}, G.}, \bibinfo{author}{{Hu}, L.},
  \bibinfo{author}{{Hunkeler}, J.S.}, \bibinfo{author}{{Ivezi{\'c}}, {\v Z}.},
  \bibinfo{author}{{Jain}, A.}, \bibinfo{author}{{Jenness}, T.},
  \bibinfo{author}{{Kanarek}, G.}, \bibinfo{author}{{Kendrew}, S.},
  \bibinfo{author}{{Kern}, N.S.}, \bibinfo{author}{{Kerzendorf}, W.E.},
  \bibinfo{author}{{Khvalko}, A.}, \bibinfo{author}{{King}, J.},
  \bibinfo{author}{{Kirkby}, D.}, \bibinfo{author}{{Kulkarni}, A.M.},
  \bibinfo{author}{{Kumar}, A.}, \bibinfo{author}{{Lee}, A.},
  \bibinfo{author}{{Lenz}, D.}, \bibinfo{author}{{Littlefair}, S.P.},
  \bibinfo{author}{{Ma}, Z.}, \bibinfo{author}{{Macleod}, D.M.},
  \bibinfo{author}{{Mastropietro}, M.}, \bibinfo{author}{{McCully}, C.},
  \bibinfo{author}{{Montagnac}, S.}, \bibinfo{author}{{Morris}, B.M.},
  \bibinfo{author}{{Mueller}, M.}, \bibinfo{author}{{Mumford}, S.J.},
  \bibinfo{author}{{Muna}, D.}, \bibinfo{author}{{Murphy}, N.A.},
  \bibinfo{author}{{Nelson}, S.}, \bibinfo{author}{{Nguyen}, G.H.},
  \bibinfo{author}{{Ninan}, J.P.}, \bibinfo{author}{{N{\"o}the}, M.},
  \bibinfo{author}{{Ogaz}, S.}, \bibinfo{author}{{Oh}, S.},
  \bibinfo{author}{{Parejko}, J.K.}, \bibinfo{author}{{Parley}, N.},
  \bibinfo{author}{{Pascual}, S.}, \bibinfo{author}{{Patil}, R.},
  \bibinfo{author}{{Patil}, A.A.}, \bibinfo{author}{{Plunkett}, A.L.},
  \bibinfo{author}{{Prochaska}, J.X.}, \bibinfo{author}{{Rastogi}, T.},
  \bibinfo{author}{{Reddy Janga}, V.}, \bibinfo{author}{{Sabater}, J.},
  \bibinfo{author}{{Sakurikar}, P.}, \bibinfo{author}{{Seifert}, M.},
  \bibinfo{author}{{Sherbert}, L.E.}, \bibinfo{author}{{Sherwood-Taylor}, H.},
  \bibinfo{author}{{Shih}, A.Y.}, \bibinfo{author}{{Sick}, J.},
  \bibinfo{author}{{Silbiger}, M.T.}, \bibinfo{author}{{Singanamalla}, S.},
  \bibinfo{author}{{Singer}, L.P.}, \bibinfo{author}{{Sladen}, P.H.},
  \bibinfo{author}{{Sooley}, K.A.}, \bibinfo{author}{{Sornarajah}, S.},
  \bibinfo{author}{{Streicher}, O.}, \bibinfo{author}{{Teuben}, P.},
  \bibinfo{author}{{Thomas}, S.W.}, \bibinfo{author}{{Tremblay}, G.R.},
  \bibinfo{author}{{Turner}, J.E.H.}, \bibinfo{author}{{Terr{\'o}n}, V.},
  \bibinfo{author}{{van Kerkwijk}, M.H.}, \bibinfo{author}{{de la Vega}, A.},
  \bibinfo{author}{{Watkins}, L.L.}, \bibinfo{author}{{Weaver}, B.A.},
  \bibinfo{author}{{Whitmore}, J.B.}, \bibinfo{author}{{Woillez}, J.},
  \bibinfo{author}{{Zabalza}, V.}, \bibinfo{year}{2018}.
\newblock \bibinfo{title}{{The Astropy Project: Building an inclusive,
  open-science project and status of the v2.0 core package}}.
\newblock \bibinfo{journal}{ArXiv e-prints} \eprint{1801.02634}.
\bibitem[{{The Astropy Collaboration} et~al.(2013){The Astropy Collaboration},
  Robitaille, Tollerud, Greenfield, Droettboom, Bray, Aldcroft, Davis,
  Ginsburg, Price-Whelan, Kerzendorf, Conley, Crighton, Barbary, Muna,
  Ferguson, Grollier, Parikh, Nair, G{\"{u}}nther, Deil, Woillez, Conseil,
  Kramer, Turner, Singer, Fox, Weaver, Zabalza, Edwards, Bostroem, Burke,
  Casey, Crawford, Dencheva, Ely, Jenness, Labrie, Lim, Pierfederici, Pontzen,
  Ptak, Refsdal, Servillat and Streicher}]{TheAstropyCollaboration2013}
\bibinfo{author}{{The Astropy Collaboration}, A.}, \bibinfo{author}{Robitaille,
  T.P.}, \bibinfo{author}{Tollerud, E.J.}, \bibinfo{author}{Greenfield, P.},
  \bibinfo{author}{Droettboom, M.}, \bibinfo{author}{Bray, E.},
  \bibinfo{author}{Aldcroft, T.}, \bibinfo{author}{Davis, M.},
  \bibinfo{author}{Ginsburg, A.}, \bibinfo{author}{Price-Whelan, A.M.},
  \bibinfo{author}{Kerzendorf, W.E.}, \bibinfo{author}{Conley, A.},
  \bibinfo{author}{Crighton, N.}, \bibinfo{author}{Barbary, K.},
  \bibinfo{author}{Muna, D.}, \bibinfo{author}{Ferguson, H.},
  \bibinfo{author}{Grollier, F.}, \bibinfo{author}{Parikh, M.M.},
  \bibinfo{author}{Nair, P.H.}, \bibinfo{author}{G{\"{u}}nther, H.M.},
  \bibinfo{author}{Deil, C.}, \bibinfo{author}{Woillez, J.},
  \bibinfo{author}{Conseil, S.}, \bibinfo{author}{Kramer, R.},
  \bibinfo{author}{Turner, J.E.H.}, \bibinfo{author}{Singer, L.},
  \bibinfo{author}{Fox, R.}, \bibinfo{author}{Weaver, B.A.},
  \bibinfo{author}{Zabalza, V.}, \bibinfo{author}{Edwards, Z.I.},
  \bibinfo{author}{Bostroem, K.A.}, \bibinfo{author}{Burke, D.J.},
  \bibinfo{author}{Casey, A.R.}, \bibinfo{author}{Crawford, S.M.},
  \bibinfo{author}{Dencheva, N.}, \bibinfo{author}{Ely, J.},
  \bibinfo{author}{Jenness, T.}, \bibinfo{author}{Labrie, K.},
  \bibinfo{author}{Lim, P.L.}, \bibinfo{author}{Pierfederici, F.},
  \bibinfo{author}{Pontzen, A.}, \bibinfo{author}{Ptak, A.},
  \bibinfo{author}{Refsdal, B.}, \bibinfo{author}{Servillat, M.},
  \bibinfo{author}{Streicher, O.}, \bibinfo{year}{2013}.
\newblock \bibinfo{title}{{Astropy: A Community Python Package for Astronomy}}.
\newblock \bibinfo{journal}{Astronomy {\&} Astrophysics, Volume 558, id.A33, 9
  pp.} \bibinfo{volume}{558}.
\newblock \eprint{1307.6212}.
\bibitem[{{Thompson} et~al.(2001){Thompson}, {Moran} and {Swenson}}]{TMS}
\bibinfo{author}{{Thompson}, A.R.}, \bibinfo{author}{{Moran}, J.M.},
  \bibinfo{author}{{Swenson}, Jr., G.W.}, \bibinfo{year}{2001}.
\newblock \bibinfo{title}{{Interferometry and Synthesis in Radio Astronomy, 2nd
  Edition}}.
\bibitem[{{Tingay} et~al.(2013){Tingay}, {Goeke}, {Bowman} et~al.}]{Tingay13}
\bibinfo{author}{{Tingay}, S.J.}, \bibinfo{author}{{Goeke}, R.},
  \bibinfo{author}{{Bowman}, J.D.}, et~al., \bibinfo{year}{2013}.
\newblock \bibinfo{title}{{The Murchison Widefield Array: The Square Kilometre
  Array Precursor at Low Radio Frequencies}}.
\newblock \bibinfo{journal}{PASA} \bibinfo{volume}{30}, \bibinfo{pages}{7}.
\newblock \eprint{1206.6945}.
\bibitem[{{Tody}(1986)}]{1986SPIE..627..733T}
\bibinfo{author}{{Tody}, D.}, \bibinfo{year}{1986}.
\newblock \bibinfo{title}{{The IRAF Data Reduction and Analysis System}}, in:
  \bibinfo{editor}{{Crawford}, D.L.} (Ed.), \bibinfo{booktitle}{Instrumentation
  in astronomy VI}, p. \bibinfo{pages}{733}.
\bibitem[{{Tody}(1993)}]{1993ASPC...52..173T}
\bibinfo{author}{{Tody}, D.}, \bibinfo{year}{1993}.
\newblock \bibinfo{title}{{IRAF in the Nineties}}, in:
  \bibinfo{editor}{{Hanisch}, R.J.}, \bibinfo{editor}{{Brissenden}, R.J.V.},
  \bibinfo{editor}{{Barnes}, J.} (Eds.), \bibinfo{booktitle}{Astronomical Data
  Analysis Software and Systems II}, p. \bibinfo{pages}{173}.
\bibitem[{{van Haarlem} et~al.(2013){van Haarlem}, {Wise}, {Gunst}, {Heald},
  {McKean}, {Hessels}, {de Bruyn}, {Nijboer}, {Swinbank} and {et
  al}}]{vanHaarlem}
\bibinfo{author}{{van Haarlem}, M.P.}, \bibinfo{author}{{Wise}, M.W.},
  \bibinfo{author}{{Gunst}, A.W.}, \bibinfo{author}{{Heald}, G.},
  \bibinfo{author}{{McKean}, J.P.}, \bibinfo{author}{{Hessels}, J.W.T.},
  \bibinfo{author}{{de Bruyn}, A.G.}, \bibinfo{author}{{Nijboer}, R.},
  \bibinfo{author}{{Swinbank}, J.}, \bibinfo{author}{{et al}},
  \bibinfo{year}{2013}.
\newblock \bibinfo{title}{{LOFAR: The LOw-Frequency ARray}}.
\newblock \bibinfo{journal}{AAP} \bibinfo{volume}{556}, \bibinfo{pages}{A2}.
\newblock \eprint{1305.3550}.
\bibitem[{van~der Walt et~al.(2011)van~der Walt, Colbert and
  Varoquaux}]{vaderwalt_numpy_2011}
\bibinfo{author}{van~der Walt, S.}, \bibinfo{author}{Colbert, S.C.},
  \bibinfo{author}{Varoquaux, G.}, \bibinfo{year}{2011}.
\newblock \bibinfo{title}{The numpy array: A structure for efficient numerical
  computation}.
\newblock \bibinfo{journal}{Computing in Science Engineering}
  \bibinfo{volume}{13}, \bibinfo{pages}{22--30}.
\bibitem[{{Wayth} et~al.(2015){Wayth}, {Lenc}, {Bell}, {Callingham},
  {Dwarakanath}, {Franzen}, {For}, {Gaensler}, {Hancock}, {Hindson},
  {Hurley-Walker}, {Jackson}, {Johnston-Hollitt}, {Kapi{\'n}ska}, {McKinley},
  {Morgan}, {Offringa}, {Procopio}, {Staveley-Smith}, {Wu}, {Zheng}, {Trott},
  {Bernardi}, {Bowman}, {Briggs}, {Cappallo}, {Corey}, {Deshpande}, {Emrich},
  {Goeke}, {Greenhill}, {Hazelton}, {Kaplan}, {Kasper}, {Kratzenberg},
  {Lonsdale}, {Lynch}, {McWhirter}, {Mitchell}, {Morales}, {Morgan}, {Oberoi},
  {Ord}, {Prabu}, {Rogers}, {Roshi}, {Shankar}, {Srivani}, {Subrahmanyan},
  {Tingay}, {Waterson}, {Webster}, {Whitney}, {Williams} and
  {Williams}}]{2015PASA...32...25W}
\bibinfo{author}{{Wayth}, R.B.}, \bibinfo{author}{{Lenc}, E.},
  \bibinfo{author}{{Bell}, M.E.}, \bibinfo{author}{{Callingham}, J.R.},
  \bibinfo{author}{{Dwarakanath}, K.S.}, \bibinfo{author}{{Franzen}, T.M.O.},
  \bibinfo{author}{{For}, B.Q.}, \bibinfo{author}{{Gaensler}, B.},
  \bibinfo{author}{{Hancock}, P.}, \bibinfo{author}{{Hindson}, L.},
  \bibinfo{author}{{Hurley-Walker}, N.}, \bibinfo{author}{{Jackson}, C.A.},
  \bibinfo{author}{{Johnston-Hollitt}, M.}, \bibinfo{author}{{Kapi{\'n}ska},
  A.D.}, \bibinfo{author}{{McKinley}, B.}, \bibinfo{author}{{Morgan}, J.},
  \bibinfo{author}{{Offringa}, A.R.}, \bibinfo{author}{{Procopio}, P.},
  \bibinfo{author}{{Staveley-Smith}, L.}, \bibinfo{author}{{Wu}, C.},
  \bibinfo{author}{{Zheng}, Q.}, \bibinfo{author}{{Trott}, C.M.},
  \bibinfo{author}{{Bernardi}, G.}, \bibinfo{author}{{Bowman}, J.D.},
  \bibinfo{author}{{Briggs}, F.}, \bibinfo{author}{{Cappallo}, R.J.},
  \bibinfo{author}{{Corey}, B.E.}, \bibinfo{author}{{Deshpande}, A.A.},
  \bibinfo{author}{{Emrich}, D.}, \bibinfo{author}{{Goeke}, R.},
  \bibinfo{author}{{Greenhill}, L.J.}, \bibinfo{author}{{Hazelton}, B.J.},
  \bibinfo{author}{{Kaplan}, D.L.}, \bibinfo{author}{{Kasper}, J.C.},
  \bibinfo{author}{{Kratzenberg}, E.}, \bibinfo{author}{{Lonsdale}, C.J.},
  \bibinfo{author}{{Lynch}, M.J.}, \bibinfo{author}{{McWhirter}, S.R.},
  \bibinfo{author}{{Mitchell}, D.A.}, \bibinfo{author}{{Morales}, M.F.},
  \bibinfo{author}{{Morgan}, E.}, \bibinfo{author}{{Oberoi}, D.},
  \bibinfo{author}{{Ord}, S.M.}, \bibinfo{author}{{Prabu}, T.},
  \bibinfo{author}{{Rogers}, A.E.E.}, \bibinfo{author}{{Roshi}, A.},
  \bibinfo{author}{{Shankar}, N.U.}, \bibinfo{author}{{Srivani}, K.S.},
  \bibinfo{author}{{Subrahmanyan}, R.}, \bibinfo{author}{{Tingay}, S.J.},
  \bibinfo{author}{{Waterson}, M.}, \bibinfo{author}{{Webster}, R.L.},
  \bibinfo{author}{{Whitney}, A.R.}, \bibinfo{author}{{Williams}, A.},
  \bibinfo{author}{{Williams}, C.L.}, \bibinfo{year}{2015}.
\newblock \bibinfo{title}{{GLEAM: The GaLactic and Extragalactic All-Sky MWA
  Survey}}.
\newblock \bibinfo{journal}{\pasa} \bibinfo{volume}{32}, \bibinfo{pages}{e025}.
\newblock \eprint{1505.06041}.

\end{thebibliography}







\end{document}